\documentclass[preprintnumbers,nofootinbib,superscriptaddress,10pt]{revtex4}
\usepackage[dvipdfmx]{graphicx} 
\usepackage{amsmath,amssymb,amsfonts,bm,cancel} 

\def\a{\alpha}  \def\g{\gamma}  \def\d{\delta}          \def\m{\mu} \def\n{\nu}     \def\r{\rho}   \def\t{\tau}       

\def\dg{\dagger}  \def\nn{\nonumber}

\newcommand{\lsp}{ \left ( } \newcommand{\rsp}{ \right ) }   \newcommand{\To}{\Rightarrow}

\renewcommand{\Im}{{\rm Im}\,}

\newcommand{\Diag}[3]{ \begin{pmatrix} #1 & 0 & 0 \\ 0 & #2 & 0 \\ 0 & 0 & #3 \\\end{pmatrix}}

\usepackage{color}

\begin{document}

\title{\large  Rephasing invariant structure of Dirac CP phase and \\
basis independent reduction of unitarity constraints for mixing matrices}

\preprint{STUPP-25-294}

\author{Masaki J. S. Yang}
\email{mjsyang@mail.saitama-u.ac.jp}
\affiliation{Department of Physics, Saitama University, 
Shimo-okubo, Sakura-ku, Saitama, 338-8570, Japan}
\affiliation{Department of Physics, Graduate School of Engineering Science,
Yokohama National University, Yokohama, 240-8501, Japan}



\begin{abstract} 

In this paper, we explore rephasing invariant structures of the Dirac CP phase $\delta$ under an approximation $U^{e}_{13} = 0$, where the 1-3 element of the diagonalization of charged leptons $U^{e}$ is neglected. 
With the further simplified condition $U^{e}_{12} = 0$, the Dirac phase reduces to a compact form 
$\delta =  \delta^{\nu} + \arg [ (U_{33}^e / U_{23}^{e}) - ( U_{33}^{\n} /  U_{23}^{\n}) ] 
- \arg [ ( U_{33}^e / U_{23}^{e}) + (U_{23}^{\n *} / U_{33}^{\n*} ) ]  $, 
and the CP phase for finite $U_{12}^{e}$ can be  understood as a generalization of this compact form. 
These results encompass almost all perturbative calculations of the CP phases
in quark and lepton mixing matrices with hierarchical masses of charged fermions, 
and are independent of any specific parametrization.
As a second result of this work, we derive a basis independent reduction of the unitarity constraints for an arbitrary unitary matrix $V$ by eliminating the elements $V_{21}, V_{22}, V_{31}, V_{32}$ using the inversion formula.
Applying the explicit rephasing transformation to this reduction yields a rephasing invariant representation of the PDG parametrization, which allows the translation of theoretical results expressed in the PDG parametrization directly into rephasing invariants.

\end{abstract} 

\maketitle

\section{Introduction}

CP violation is one of the most fundamental problems in particle physics and stimulates both theoretical and experimental research in flavor physics. 
The measurement of the Dirac CP phase in the lepton mixing matrix is currently ongoing \cite{NOvA:2021nfi, T2K:2021xwb}, and significant improvements in precision are expected from future experiments \cite{DUNE:2020jqi, Hyper-KamiokandeProto-:2015xww}.
In addition, indirect searches for Majorana phases continue through neutrinoless double beta decay experiments \cite{KamLAND-Zen:2022tow}.

On the theoretical side, general studies of the Dirac phase have been performed extensively 
  \cite{Xing:2002sw, Antusch:2005kw, Kuo:2005pf, Farzan:2006vj, Hochmuth:2007wq, Ge:2011qn, Marzocca:2013cr,  Petcov:2014laa, Dasgupta:2014ula, Chiu:2015ega}. 
For the Majorana phases as well, correlations with other CP-violating observables have been discussed in detail in studies of generalized CP symmetries 
\cite{Ecker:1981wv, Ecker:1983hz, Gronau:1985sp, Ecker:1987qp, Neufeld:1987wa, Ferreira:2009wh, Feruglio:2012cw, Holthausen:2012dk, Ding:2013bpa, Girardi:2013sza, Nishi:2013jqa, Feruglio:2013hia, Chen:2014wxa, Ding:2014ora, Chen:2014tpa, Li:2015jxa, Turner:2015uta, Rodejohann:2017lre, Penedo:2017vtf, Nath:2018fvw, Yang:2021smh, Yang:2021xob, Ohki:2023zsn}, including the $\mu$-$\tau$ reflection symmetry \cite{Harrison:2002et, Grimus:2003yn, Grimus:2005jk, Joshipura:2007sf, Adhikary:2009kz, Joshipura:2009tg, Xing:2010ez, Ge:2010js, He:2011kn, Gupta:2011ct, He:2012yt, Joshipura:2015dsa, Xing:2015fdg, Chen:2015siy, Samanta:2017kce, Nishi:2018vlz, Sinha:2018xof, Huang:2018fog, Pan:2019qcc, Chakraborty:2019rjc, Liao:2019qbb, Yang:2020qsa, Zhao:2021dwc, Bao:2022kon, Kumar:2024zfb}.

In recent developments, through the study of rephasing invariants with the determinant of the mixing matrix \cite{Yang:2025hex,Yang:2025cya,Yang:2025law,Yang:2025ftl,Yang:2025dhm,Yang:2025vrs}, 
an explicit rephasing transformation to the PDG parametrization was derived for an arbitrary mixing matrix \cite{Yang:2025dkm}.
Subsequent studies applied the explicit transformation to fermion diagonalization matrices, clarifying the rephasing structure of CP phases across different parametrizations \cite{Yang:2025qlg, Yang:2026ulu, Yang:2026wjg}. 
For more than four decades since the original work of Chau and Keung \cite{Chau:1984fp}, this rephasing procedure has been employed only at a conceptual level.

In this paper, we explore such rephasing structures of the Dirac CP phase $\d$ in the PDG parametrization.
By imposing the approximation $U_{13}^{e} = 0$ for diagonalization matrix of charged leptons $U^{e}$, 
 the phase $\delta$ becomes a function of the Dirac-like phase of neutrinos $\delta^{\nu}$ and two relative phases $\rho_{1}-\rho_{2}$ and $\rho_{2}-\rho_{3}$. 
To express the nontrivial arguments of neutrinos $U_{21}^{\nu}$ and $U_{22}^{\nu}$ in terms of these three phases, we present a decomposition of an arbitrary unitary matrix into a rephasing invariant form of PDG parametrization and the explicit rephasing transformation.

The paper is organized as follows.
In the next section, we review the explicit rephasing transformation to the PDG parametrization and derive expressions for the Dirac and Majorana phases.
In Sec. III, we investigate the rephasing invariant structure of the Dirac CP phase and 
demonstrate how an arbitrary unitary matrix can be decomposed in the rephasing invariant representation of the PDG parametrization and the explicit rephasing transformation. 
The final section is devoted to the summary.

\section{Explicit rephasing transformation to PDG parametrization}

Here, we review the explicit rephasing transformation through the derivation of the Dirac phase $\delta$ and the Majorana phases $\alpha_{2,3}$ \cite{Yang:2025dkm}.
The standard PDG parametrization of the lepton mixing matrix is given by \cite{ParticleDataGroup:2018ovx} 
\begin{align}
U^{\rm PDG} &= U^{0} P \equiv
\begin{pmatrix}
c_{12} c_{13} & s_{12} c_{13} & s_{13} e^{-i\d} \\
-s_{12} c_{23} - c_{12} s_{23} s_{13}  e^{i \d} & c_{12} c_{23} - s_{12} s_{23} s_{13} e^{i \d} & s_{23} c_{13} \\
s_{12} s_{23} - c_{12} c_{23} s_{13} e^{i \d} & -c_{12} s_{23} - s_{12} c_{23} s_{13} e^{ i \d} & c_{23} c_{13}
\end{pmatrix} \nn \\
& \times {\rm diag} (1 ,  e^{ i \a_{2} / 2} , e^{ i \a_{3} / 2}) \, .
\label{PDG}
\end{align}
Suppose that the standard form $U^{0}$ is mapped to a mixing matrix $U$ in an arbitrary basis through a rephasing transformation 
\begin{align}
\begin{pmatrix}
U_{e1} & U_{e2} & U_{e3}  \\
U_{\m1} & U_{\m 2} & U_{\m 3}  \\
U_{\t 1} & U_{\t 2} & U_{\t 3}  \\
\end{pmatrix}
= 
\Diag{e^{i \g_{L1}}}{e^{i \g_{L2}}}{e^{i \g_{L3}}}
\begin{pmatrix}
|U_{e1}| & |U_{e2}| & |U_{e3}| e^{-i\d} \\
U_{\m1}^{0} & U_{\m 2}^{0} & |U_{\m 3}|  \\
U_{\t 1}^{0} & U_{\t 2}^{0} & |U_{\t 3}|  \\
\end{pmatrix}
\Diag{e^{ - i \g_{R1}}}{e^{ - i \g_{R2}}}{e^{ - i \g_{R3}}} . 
\end{align}
Focusing on $U_{e3} = |U_{e3}| e^{-i(\delta - \gamma_{L1} + \gamma_{R3})}$,
the phase factor $e^{i\delta}$ is determined by
\begin{align}
U_{e3}^{*} = |U_{e3}| e^{i(\d - \g_{L1} + \g_{R3} )}  \, , ~~ 
|U_{e3}| e^{i \d } = U_{e3}^{*}e^{i(\g_{L1} - \g_{R3} )}  \, . 
\end{align}
Eliminating the remaining four phases using the determinant and matrix elements with trivial phases, the Dirac phase $\delta$ is solved as
\begin{align}
\arg \left[ {U_{e1} U_{e2} U_{\m3} U_{\t 3} } \right]
 & = 2 \g_{L1} - \g_{R1}  - \g_{R2}  + \g_{L2} + \g_{L3} - 2 \g_{R3}  \, , \nn \\
 \arg \left[ {U_{e1} U_{e2} U_{\m3} U_{\t 3} \over \det U } \right] & = 
 \g_{L1}  -  \g_{R3} 
 ~~ \To ~~ 
\d = \arg \left[ {U_{e1} U_{e2} U_{\m3} U_{\t 3}   \over U_{e3} \det U } \right] \, . 
\label{Dirac}
\end{align}

Another phase difference $\gamma_{R1} - \gamma_{R2}$ 
corresponds to the well-known result of the  Majorana phase \cite{Doi:1980yb} 
\begin{align}
 \arg [ U_{e1}^{*} U_{e2}] & = \g_{R1}  - \g_{R2} = \a_{2} / 2 \, ,
\end{align}
because the parameter $\a_{2}$ is defined in the basis where $\gamma_{R1} = 0$. 
Moreover, by combining with $\delta$, one can obtain an explicit expression for $\alpha_{3}/2$; 
\begin{align}
\a_{3} / 2 = \arg [ U_{e1}^{*} U_{e3}]  +  \arg \left[ {U_{e1} U_{e2} U_{\m3} U_{\t 3}   \over U_{e3} \det U } \right] 
=  \arg \left[ {U_{e2} U_{\m3} U_{\t 3}   \over  \det U } \right] \, . 
\end{align}

This observation naturally motivates us to express the left-handed phases $\gamma_{Li}$ in terms of the arguments of the matrix elements. 
The conditions for this purpose are
\begin{align}
\g_{L1} - \g_{R1} = \arg U_{e1} \, , ~~
\g_{L1} - \g_{R2} = \arg U_{e2} \, , ~~
\g_{L2} - \g_{R3} = \arg U_{\m3} \, , ~~
\g_{L3} - \g_{R3} = \arg U_{\t 3} \, .
\end{align}
Reflecting the freedom of the overall phase, 
one of the six phases $\gamma_{Li}$ and $\gamma_{Ri}$ remains undetermined. 
To preserve the structure of the Majorana phases, we leave $\gamma_{R1}$ as an undetermined variable.
Since it is eventually canceled by the left- and right-handed phase matrices, 
the transformation of $U$ into the PDG parametrization is found to be
\begin{align}
U = 
\Diag{e^{i \arg U_{e1} } }{e^{ i  \arg \left[ {\det U \over  U_{e2}  U_{\t 3} } \right]}}{e^{ i \arg \left[ {  \det U \over U_{e2} U_{\m3} } \right] }}
U^{0}
\Diag{1}{e^{ i \arg [{U_{e2}\over U_{e1} }] } }{e^{  i \arg \left[ { U_{e2} U_{\m3} U_{\t 3} \over \det U } \right] }} . 
\end{align}
This provides an explicit rephasing transformation to the PDG parametrization, which had remained a largely conceptual operation for almost half a century. 
By combining the left- and right-handed phases, cancellations occur in an alternating way, yielding a sum of the five remaining variant phases equal to 
\begin{align}
 \arg U_{e1} + \arg \left[{U_{e2} \over  U_{e1}} \right] 
+ \arg \left[ { \det U \over U_{e2}  U_{\t 3} } \right] 
 + \arg \left[ { U_{e2} U_{\m3} U_{\t 3} \over \det U } \right] 
 + \arg \left[ {  \det U \over U_{e2} U_{\m3} } \right] 
=  \arg\det U \, . 
\end{align}

With this explicit rephasing, the independent six phases of a unitary matrix can be expressed in terms of the arguments of its matrix elements. By explicitly performing the matrix multiplication, one finds 
\begin{align}
U  = 
\begin{pmatrix}
U_{e1} & U_{e2} & e^{i \arg [ { U_{e1}U_{e2} U_{\m3} U_{\t 3} \over \det U }] } |U_{e3}| e^{-i\d} \\
e^{- i  \arg \left[ { U_{e2}  U_{\t 3} \over \det U } \right]} U_{\m1}^{0} & e^{- i  \arg \left[ { U_{e1} U_{\t 3} \over \det U } \right] } U_{\m 2}^{0} & U_{\m 3}  \\
e^{- i \arg \left[ { U_{e2} U_{\m3} \over  \det U } \right] } U_{\t 1}^{0} & e^{-i \arg \left[ { U_{e1} U_{\m3} \over  \det U } \right] } U_{\t 2}^{0} &  U_{\t 3}  \\
\end{pmatrix} \, . 
\end{align}
Consequently, the four nontrivial arguments are written in terms of other rephasing invariants 
\begin{align}
\arg U_{\m1}^{0} & = \arg \left[ { U_{e2} U_{\m1} U_{\t 3} / \det U } \right] \, , ~~ 
\arg U_{\m 2}^{0} = \arg \left[ { U_{e1} U_{\m2} U_{\t 3} / \det U } \right] \, , \\
\arg U_{\t 1}^{0} & = \arg \left[ { U_{e2} U_{\m3} U_{\t1} /  \det U } \right] \, , ~~ 
\arg U_{\t 2}^{0}  = \arg \left[ { U_{e1} U_{\m3} U_{\t2} /  \det U } \right] \, . 
\end{align}
Since the elements $U_{e1}, U_{e2}, U_{\mu 3},U_{\tau 3}$ and $\det U$ have only trivial arguments in the PDG parametrization, the validity of these expressions is manifest.

This explicit rephasing can be applied not only to the lepton mixing matrices, 
but also to the neutrino and charged-lepton diagonalization matrices $U^{\nu}$ and $U^{e}$.
We define their explicit rephasing transformation $U^{\nu,e} = \Phi^{\nu,e}_{L} \, U^{\nu,e 0} \, \Phi^{\nu,e}_{R}$,
where $U^{\nu,e 0}$ denotes the PDG standard form and $\Phi^{\nu,e}_{L,R}$ are phase matrices expressed in terms of the corresponding matrix elements.
The  Dirac-like phases $\delta^{\nu,e}$ are also written in the rephasing invariants 
\begin{align}
\d^{\n , e} =  \arg \left[ {U_{11}^{\n , e} U_{12}^{\n , e} U_{23}^{\n , e} U_{3 3}^{\n , e}   \over U_{13}^{\n , e} \det U^{\n , e} } \right] \, . 
\end{align}

From this, the lepton mixing matrix $U \equiv U^{e \dagger} U^{\nu}$ is given by
\begin{align}
U = \Phi^{e \dagger}_{R} \, U^{e 0\dagger} \, \Phi_L^{e\dagger} \Phi_L^\nu \, U^{\nu 0} \, \Phi_R^{\n} \, .
\end{align}
Among these sources of CP violation, only the left-handed phases of each fermion contribute to the Dirac phase, while the right-handed phases $\Phi_R^{\n}$ affect the Majorana phases. 
By combining the left-handed phases as $\Phi^L \equiv \Phi_e^{L\dagger} \Phi_\nu^L \equiv {\rm diag} \, ( e^{i\r_{1}} \, , \, e^{i\r_{2}} \, , \, e^{i \r_{3}})$, 
relative phases between $U^{\n 0}$ and $U^{e 0}$ are 
\begin{align}
\r_{1} = \arg \left[ { U^{e *}_{11} U^{\n}_{11} } \right] , ~~
\r_{2} =  \arg \left[ { \det U^{e *} \over U^{e *}_{12}  U^{e *}_{33} } { \det U^{\n} \over U^{\n}_{12}  U^{\n}_{33} } \right] , ~~ 
\r_{3} = \arg \left[ { \det U^{e *} \over  U^{e *}_{12} U^{e *}_{23} } { \det U^{\n} \over U^{\n}_{12} U^{\n}_{23} } \right] . 
\end{align}
Furthermore, only two independent combinations of the relative phases $\r_{i} - \r_{j}$ are physically relevant 
\begin{align}
\r_{1} -\r_{2}  & =  \arg \left[  { U^{e *}_{11}  U^{e *}_{12}  U^{e *}_{33} \over \det U^{e *}  } { U^{\n}_{11} U^{\n}_{12}  U^{\n}_{33} \over \det U^{\n} } \right] ,  ~~
 \r_{2} - \r_{3} =  \arg \left[ { U^{e *}_{23} U^{\n}_{23}   \over U^{e*}_{33} U^{\n}_{33}  } \right]  ,  \\
 \r_{3} - \r_{1}  & = \arg \left[ {  \det U^{e *} \over U^{e *}_{11} U^{e *}_{12} U^{e *}_{23} } {\det U^{\n} \over U^{\n}_{11} U^{\n}_{12} U^{\n}_{23}  } \right]  . 
\end{align}
In this way, the relative phases between the fermion diagonalization matrices are indeed rephasing invariants under the transformation property of $U_{\rm MNS} = U_{e}^{\dg} U_{\n}$. 
The Dirac phase $\d$ is a function of four phases, the Dirac-like phases $\d_{\n}$ and $\d_{e}$, together with two independent relative phases.

\section{Rephasing invariant structure of Dirac CP Phase}

In this section, we investigate the behavior of the Dirac CP phase by rephasing invariants.
In a completely general setting, the analysis becomes cumbersome and does not lead to particularly transparent results.
We therefore introduce an approximation in order to obtain a more compact expression for the CP phase.

\begin{description}
\item[\bf Approximation:] 
In the lepton mixing matrix $U = U^{e \dg} U^{\n}$, the 1-3 element of $U^{e}$ is negligible, $U_{13}^{e} = 0$.

\item[\bf Justification:] 
When the charged-lepton mass matrix $m_{e}$ possesses chiral symmetries of the first and second generations,
$m_{e} = D_{L} m_{e} D_{R}$, all corresponding singular values and mixings vanish. 
Here, $D_{L,R} \equiv {\rm diag} (e^{i\phi_{L,R}^{1} } , e^{i \phi_{L,R}^{2}} , 1)$ and $\phi_{L,R}^{1,2}$ denote phase parameters.
These chiral symmetries are only approximate in reality,  
and the mixing angles are suppressed by $m_{ei} / m_{ej}$, powers of the ratios of singular values.

\end{description}

Before examining the general behavior in this situation, 
let us consider a further simplified approximation $U_{13}^{e} = 0$ and $U_{12}^{e} = 0$.
As in the previous works \cite{Yang:2025dkm, Yang:2025qlg, Yang:2026ulu, Yang:2026wjg}, 
by applying a partial inversion of the diagonalization matrix $U^{e}$, 
the mixing matrix becomes
\begin{align}
U &= 
\begin{pmatrix}
U_{11}^{e *} & 0 & 0 \\[2pt]
0 & {U_{11}^e U_{33}^e \over \det U^{e} } & - {U_{11}^{e} U_{23}^{e} \over \det U^{e } } \\[2pt]
0 & U_{23}^{e *} & U_{33}^{e *} \\
\end{pmatrix} 
\begin{pmatrix}
U_{11}^{\n} & U_{12}^{\n} & U_{13}^{\n} \\[2pt]
U_{21}^{\n} & U_{22}^{\n}  & U_{23}^{\n} \\[2pt]
U_{31}^{\n} & U_{32}^{\n} & U_{33}^{\n} \\
\end{pmatrix} 
 = 
\begin{pmatrix}
U_{11}^{e *} U_{11}^{\n} &U_{11}^{e *}  U_{12}^{\n} & U_{11}^{e *} U_{13}^{\n}  \\[2pt]
*  &  * &   {U_{11}^e \over \det U^{e} }  (U_{33}^e U_{23}^{\n} -  U_{23}^{e} U_{33}^{\n})   \\[2pt]
* & * &U_{23}^{e *} U_{23}^{\n} + U_{33}^{e *}U_{33}^{\n}
\end{pmatrix} . 
\end{align}
The matrix elements denoted by $*$ are irrelevant to the phase evaluation, 
and the CP phase is found to be
\begin{align}
\d & = \arg \left[ { U_{11}^{\n}  U_{12}^{\n} \over U_{13}^{\n}  \det U^{\n}} 
{ U_{33}^e U_{23}^{\n} -  U_{23}^{e} U_{33}^{\n} \over  U_{23}^{e} U_{23}^{\n *} + U_{33}^{e} U_{33}^{\n*} }  \right] 
 = \d^{\n} + 
\arg \left[  1 -  {U_{23}^{e} U_{33}^{\n} \over U_{33}^e U_{23}^{\n} }   \right] 
- \arg \left[  1 + {U_{23}^{e} U_{23}^{\n *} \over U_{33}^{e} U_{33}^{\n*} } \right]  \nn \\ 
& =    \d^{\n} + 
\arg \left[  { U_{33}^e \over U_{23}^{e}} -  { U_{33}^{\n} \over  U_{23}^{\n} }   \right] 
- \arg \left[  { U_{33}^e \over U_{23}^{e}} + { U_{23}^{\n *} \over  U_{33}^{\n*} } \right]  . 
\label{simpleCP}
\end{align}
This simplicity originates from the fact that the 2-3 mixing of charged leptons leaves the first column of $U^{\n}$ unchanged. 
By contrast, as we will confirm later, a finite $U_{12}^{e}$ leads to a more involved behavior. 
For $|U_{23}^{e} / U_{33}^{e}| \ll 1$, the perturbative expansion $\arg [1 + x] = \Im \log [1 + x] \simeq \Im x$ produces 
\begin{align}
\d & \simeq
  \d^{\n} -  \Im \left[ {U_{23}^{e} \over U_{33}^e } \lsp { U_{33}^{\n} \over  U_{23}^{\n} } + { U_{23}^{\n *} \over U_{33}^{\n*} } \rsp \right] . 
\end{align}

By choosing the standard parameterization, the mixing matrix is described as
\begin{align}
U & = 
\begin{pmatrix}
1 & 0 & 0 \\
0 & c_{23}^{e} & - s_{23}^{e} \\
0 & s_{23}^{e} & c_{23}^{e}
\end{pmatrix}
\Diag{e^{i \r_{1}}}{e^{ i \r_{2}}}{e^{ i \r_{3}}} 
\begin{pmatrix}
1 & 0 & 0 \\
0 & c_{23}^{\n} & s_{23}^{\n} \\
0 & - s_{23}^{\n} & c_{23}^{\n}
\end{pmatrix}
\begin{pmatrix}
c_{13}^{\n} & 0 & s_{13}^{\n} e^{- i \d^{\n}} \\
0 & 1 & 0 \\
- s_{13}^{\n} e^{i \d_{\n}} & 0 & c_{13}^{\n}
\end{pmatrix}
\begin{pmatrix}
c_{12}^{\n} & s_{12}^{\n} & 0 \\
- s_{12}^{\n} & c_{12}^{\n} & 0 \\
0 & 0 & 1 
\end{pmatrix} \nn \\
\label{para1}
 & =
\begin{pmatrix}
 e^{i \rho _1} c^{\nu}_{12} c^{\nu }_{13} & e^{i \rho _1} c^{\nu }_{13} s^{\nu }_{12} & s^{\nu }_{13} e^{i( \rho _1 - \delta ^{\nu })} \\
* & *  & c^{\nu }_{13} (e^{i \rho _2} c^e_{23} s^{\nu }_{23} - e^{i \rho _3} s^e_{23} c^{\nu }_{23} ) \\
 * & *  & c^{\nu }_{13} (e^{i \rho _2} s^e_{23} s^{\nu }_{23} + e^{i \rho _3} c^e_{23} c^{\nu }_{23} ) \\
\end{pmatrix} . 
\end{align}
Separating the contributions of the first row and the third column, 
the CP phase $\d$ is indeed described by the two CP phases $\d^{\n}$ and $\r_{2} - \r_{3}$, 
\begin{align}
\d & = \d^{\n} + \arg \left[ 
\frac{ e^{i (\rho _2 - \rho _3)}c^e_{23} s^{\nu }_{23} -  s^e_{23} c^{\nu }_{23} }
{ e^{ i (\r_{2} - \r_3)} c^e_{23} c^{\nu }_{23}+ s^e_{23} s^{\nu }_{23}  } \right] \, .
\end{align}
This result is consistent with the rephasing-invariant expression, and the second term depends only on the $2$-$3$ mixing and its relative phase.
Since the matrix elements $U_{12}^{e}$ and $U_{13}^{e}$ are expected to be small due to the chiral symmetries, 
the resulting expression for the Dirac phase represents a dominant contribution in a wide class of models.

The suitability of these approximations for the PDG phase can be explained as follows.
Taking the further limit $U_{23}^{\n} \to 0$, 
\begin{align}
\d & = 
\arg \left[ - { U_{11}^{\n}  U_{12}^{\n} \over U_{13}^{\n}  \det U^{\n}} 
{  U_{23}^{e} U_{33}^{\n} \over  U_{33}^{e} U_{33}^{\n*} }  \right] .
\end{align}
In this case, the mixing matrix is reduced to 
\begin{align}
U = 
\begin{pmatrix}
1 & 0 & 0 \\
0 & c_{23}^{e} & - s_{23}^{e} \\
0 & s_{23}^{e} & c_{23}^{e}
\end{pmatrix}
\Diag{e^{i \r_{1}}}{e^{ i \r_{2}}}{e^{ i \r_{3}}} 
\begin{pmatrix}
c_{13}^{\n} & 0 & s_{13}^{\n} \\
0 & 1 & 0 \\
- s_{13}^{\n} & 0 & c_{13}^{\n}
\end{pmatrix}
\begin{pmatrix}
c_{12}^{\n} & s_{12}^{\n} & 0 \\
- s_{12}^{\n} & c_{12}^{\n} & 0 \\
0 & 0 & 1 
\end{pmatrix}  . 
\label{para1}
\end{align}
From a simple field redefinition, 
the phase is written without the argument function
\begin{align}
\d = \r_{3} - \r_{2} + \pi \, . 
\end{align}
The result for finite $U_{ij}^{e}$ without the approximations 
can be regarded as a generalization of the relative phase.

\subsection{ Rephasing invariant structure of the more general CP phase}

In the CKM mixing matrix, the 1-2 element has a magnitude of $|V_{us}| \simeq 0.22$, which may be too large to treat the 1-2 mixing of charged fermions as negligible.
Consequently, it is important to evaluate the CP phase for a finite $U_{12}^{e}$. 
Within the present approximation, the most general form of the mixing matrix is written as
\begin{align}
U &= 
\begin{pmatrix}
U_{11}^{e *} & - {U_{12}^e U_{33}^e \over \det U^{e}} & { U_{12}^{e} U_{23}^{e} \over \det U^{e}}  \\[2pt]
U_{12}^{e *} & {U_{11}^e U_{33}^e \over \det U^{e} } & - {U_{11}^{e} U_{23}^{e} \over \det U^{e } } \\[2pt]
0 & U_{23}^{e *} & U_{33}^{e *} \\
\end{pmatrix} 
\begin{pmatrix}
U_{11}^{\n} & U_{12}^{\n} & U_{13}^{\n} \\[2pt]
U_{21}^{\n} & U_{22}^{\n} & U_{23}^{\n} \\[2pt]
U_{31}^{\n} & U_{32}^{\n} & U_{33}^{\n} \\
\end{pmatrix} 
 = 
\begin{pmatrix}
\tilde U_{e1} & \tilde U_{e2} & U_{11}^{e *} U_{13}^{\n} + { U_{12}^{e} \over \det U^{e}} (U_{23}^{e} U_{33}^{\n} - U_{33}^e U_{23}^{\n}) \\[2pt]
*  &  * & U_{12}^{e *}  U_{13}^{\n} +  {U_{11}^e \over \det U^{e} }  (U_{33}^e U_{23}^{\n} -  U_{23}^{e} U_{33}^{\n})   \\[2pt]
* & * & U_{23}^{e *} U_{23}^{\n} + U_{33}^{e *}U_{33}^{\n}
\end{pmatrix} , 
\end{align}
where
\begin{align}
\tilde U_{e1} & = U_{11}^{e *} U_{11}^{\n} + { U_{12}^{e} \over \det U^{e}} (U_{23}^{e} U_{31}^{\n} - U_{33}^e  U_{21}^{\n}) \, , \\
\tilde U_{e2} & = U_{11}^{e *} U_{12}^{\n} + { U_{12}^{e} \over \det U^{e}}  (U_{23}^{e} U_{32}^{\n} - U_{33}^e U_{22}^{\n} ) \, . 
\end{align}
The Dirac CP phase $\delta$ in the general situation is 
\begin{align}
\d = \arg \left[ 
{ U_{11}^{e *} U_{11}^{\n} + { U_{12}^{e} \over \det U^{e}} (U_{23}^{e} U_{31}^{\n} - U_{33}^e  U_{21}^{\n}) 
\over ( U_{11}^{e *} U_{12}^{\n} + { U_{12}^{e} \over \det U^{e}}  (U_{23}^{e} U_{32}^{\n} - U_{33}^e U_{22}^{\n} ) )^{*}}
{  U_{12}^{e *}  U_{13}^{\n} +  {U_{11}^e \over \det U^{e} }  (U_{33}^e U_{23}^{\n} -  U_{23}^{e} U_{33}^{\n})  
\over  U_{11}^{e *} U_{13}^{\n} + { U_{12}^{e} \over \det U^{e}} (U_{23}^{e} U_{33}^{\n} - U_{33}^e U_{23}^{\n}) }
{ \det U^{e} \det U^{\n *}  \over U_{23}^{e } U_{23}^{\n *} + U_{33}^{e } U_{33}^{\n *} }
 \right ] . 
 \label{generald}
\end{align}
Compared with the simplified case, the effects of a finite $U_{12}^{e}$ are more involved.

The expression can be arranged to isolate the result~(\ref{simpleCP}) under the approximation $U_{12}^{e} = 0$, 
\begin{align}
\d &= \d^{\n} + 
\arg \left[  1 -  {U_{23}^{e} U_{33}^{\n} \over U_{33}^e U_{23}^{\n} } \right] 
- \arg \left[ 1 + {U_{23}^{e} U_{23}^{\n *} \over U_{33}^{e} U_{33}^{\n*} }  \right] \nn  \\ 
& + 
 \arg \left[ 
{ 1 + { U_{12}^{e} \over U_{11}^{e *} U_{11}^{\n}  \det U^{e}} (U_{23}^{e} U_{31}^{\n} - U_{33}^e  U_{21}^{\n}) 
\over ( 1 + { U_{12}^{e} \over U_{11}^{e *} U_{12}^{\n} \det U^{e}}  (U_{23}^{e} U_{32}^{\n} - U_{33}^e U_{22}^{\n} ) )^{*}}
{ 1 +  {U_{12}^{e *}  U_{13}^{\n}  \det U^{e}  \over U_{11}^e  (U_{33}^e U_{23}^{\n} -  U_{23}^{e} U_{33}^{\n})  }  
\over  1 + { U_{12}^{e} \over U_{11}^{e *} U_{13}^{\n} \det U^{e}} (U_{23}^{e} U_{33}^{\n} - U_{33}^e U_{23}^{\n}) }
 \right ] . 
\end{align}
If the magnitude $|U_{12}^{e}/U_{11}^{e}|$ is sufficiently small, 
the expansion $\arg[1 + x] \simeq \Im x$ yields 
\begin{align}
\d &\simeq \d^{\n} + 
\arg \left[  1 -  {U_{23}^{e} U_{33}^{\n} \over U_{33}^e U_{23}^{\n} } \right] 
- \arg \left[ 1 + {U_{23}^{e} U_{23}^{\n *} \over U_{33}^{e} U_{33}^{\n*} }  \right] 
+ \Im \left [ 
 { U_{12}^{e *}  U_{13}^{\n}  \det U^{e}  \over U_{11}^e  (U_{33}^e U_{23}^{\n} -  U_{23}^{e} U_{33}^{\n})  }  
 \right ]  \nn \\ 
& + 
 \Im \left[ 
 { U_{12}^{e} \over  U_{11}^{e *} \det U^{e}} 
 \lsp {U_{23}^{e} U_{31}^{\n} - U_{33}^e  U_{21}^{\n} \over  U_{11}^{\n} }
+ { U_{23}^{e} U_{32}^{\n} - U_{33}^e U_{22}^{\n} \over  U_{12}^{\n} } 
 - {U_{23}^{e} U_{33}^{\n} - U_{33}^e U_{23}^{\n} \over  U_{13}^{\n}} \rsp 
\right ] . 
\end{align}
This expression encompasses the behavior of the Dirac CP phases in most models with chiral symmetries, 
and  does not depend on a specific parametrization due to its rephasing invariance.
When $U^{e}$ has a CKM-like structure with
$|U^{e}_{12}| \lesssim 0.2$, $|U^{e}_{23}| \lesssim 0.05$, and $|U^{e}_{13}| \lesssim 0.01$,
the observed value $|U_{e3}| \simeq 0.15$ suggests that the contribution from $U^{e}_{13}$ is at the level of  6\%.
Since this is comparable to the current experimental uncertainty, 
the approximation is justified, and a perturbative treatment of $U^{e}_{12}$ also retains sufficient accuracy.
In particular, if the magnitude of elements $|U_{12}^{e} U_{23}^{e}|$ is also as small as $\mathcal{O}(0.01)$, several additional terms can be neglected.

By replacing fermions $\nu, e \to d, u$, the same analysis can be applied to the CKM matrix itself. 
Since the up-type fermions have a more hierarchical mass spectrum, 
the chiral symmetries suggest a relation $|U_{13}^{u}| \ll |U_{13}^{d}|$ and the approximation remains valid in most cases. 
Therefore, the present result provides an almost general description of CP violation for hierarchical mass matrices of charged fermions.
Previously, representing this system required cumbersome variable transformations
and complicated combinations of trigonometric functions \cite{Yang:2024ulq, Yang:2025yst}. Such $\sin \d$-type  expressions are intricate and make it difficult to grasp the global structure of the CP phase.

\subsection{Basis independent reduction for unitarity constraints}

The above expressions involve matrix elements with nontrivial arguments $U_{21}^{\n}, U_{22}^{\n}, U_{31}^{\n}$, and $U_{32}^{\n}$. 
Due to the unitarity constraints, these elements should be functions of $\delta^{\nu}$ and rephasings.
To make this explicit, for an arbitrary unitary matrix $V$, 
substituting the inversion formula in the lower-left sub-block back into itself again,
\begin{align}
\begin{pmatrix}
 V_{21} & V_{22}  \\
 V_{31} & V_{32}  \\
\end{pmatrix}
& = {1\over \det V^{*}}
\begin{pmatrix}
 V_{13}^* V_{32}^*-V_{12}^* V_{33}^* & V_{11}^* V_{33}^*-V_{13}^* V_{31}^*  \\
 V_{12}^* V_{23}^*-V_{13}^* V_{22}^* & V_{13}^* V_{21}^*-V_{11}^* V_{23}^*  \\
\end{pmatrix}  \nn
\\ & = {1\over \det V^{*}}
\begin{pmatrix}
 V_{13}^* ({V_{13} V_{21} - V_{11} V_{23} \over \det V} ) -V_{12}^* V_{33}^* & V_{11}^* V_{33}^*-V_{13}^* ({ V_{12} V_{23} - V_{13} V_{22} \over \det V}) \\[2pt]
 V_{12}^* V_{23}^*-V_{13}^* ({ V_{11} V_{33} - V_{13} V_{31}  \over \det V})& V_{13}^* ({ V_{13} V_{32} - V_{12}  V_{33} \over \det V})-V_{11}^* V_{23}^*  \\
\end{pmatrix} \, . 
\end{align}
At this stage, the terms $|V_{13}|^{2} V_{ij}$ can be moved to the left-hand side, 
\begin{align}
(1 - |V_{13}|^{2})
\begin{pmatrix}
 V_{21} & V_{22}  \\
 V_{31} & V_{32}  \\
\end{pmatrix}
&  = {1\over \det V^{*}}
\begin{pmatrix}
 -V_{12}^* V_{33}^* & V_{11}^* V_{33}^*  \\
 V_{12}^* V_{23}^* & -V_{11}^* V_{23}^*  \\
\end{pmatrix} 
- V_{13}^{*}
\begin{pmatrix}
V_{11} V_{23}  & V_{12} V_{23}  \\
 V_{11} V_{33} & V_{12}  V_{33}  \\
\end{pmatrix} 
\, . 
\end{align}
Since the lower-left sub-block is expressed in terms of the remaining matrix elements, 
they can be eliminated by substitution, which yields the following reduction 
\begin{align}
\begin{pmatrix}
 V_{11} & V_{12} & V_{13} \\
 V_{21} & V_{22} & V_{23} \\
 V_{31} & V_{32} & V_{33} \\
\end{pmatrix}
& = 
\begin{pmatrix}
V_{11} & V_{12} & 0\\[2pt]
- {V_{12}^{*} V_{33}^{*} \det V \over 1 - |V_{13}|^{2}}  & {V_{11}^{*} V_{33}^{*} \det V \over 1 - |V_{13}|^{2}}  & V_{23} \\[2pt]
{V_{12}^{*} V_{23}^{*} \det V \over 1 - |V_{13}|^{2}}  & - {V_{11}^{*} V_{23}^{*} \det V \over 1 - |V_{13}|^{2}}  & V_{33} \\
\end{pmatrix} 
+ 
\begin{pmatrix}
 0 & 0 & V_{13} \\[2pt]
- {V_{13}^{*} V_{11} V_{23} \over 1 - |V_{13}|^{2} } &- {V_{13}^{*} V_{12} V_{23} \over 1 - |V_{13}|^{2} }   & 0 \\[2pt]
- {V_{13}^{*} V_{11} V_{33} \over 1 - |V_{13}|^{2} } & -{V_{13}^{*} V_{12} V_{33} \over 1 - |V_{13}|^{2} }    & 0 \\
\end{pmatrix} . \label{reduction}
\end{align}

Applying the explicit rephasing transformation to this reduction, 
this provides a rephasing invariant representation of the PDG parametrization~(\ref{PDG}) 
\begin{align}
V^{0} & = 
\Diag{e^{- i \arg V_{11} } }{e^{ i  \arg \left[ { V_{12}  V_{3 3} \over \det V } \right]}}{e^{ i \arg \left[ { V_{12} V_{23} \over  \det V } \right] }}
V
\Diag{1}{e^{ i \arg \left[{ V_{11} \over V_{12} } \right] } }{e^{  i \arg \left[ {  \det V \over V_{12} V_{23} V_{33} } \right] }} \nn \\
& = 
\begin{pmatrix}
|V_{11}| & |V_{12}| & 0\\[2pt]
- { |V_{12} V_{33} |\over 1 - |V_{13}|^{2}} & { |V_{11} V_{33} |\over 1 - |V_{13}|^{2}}  & |V_{23}| \\[2pt]
{ |V_{12} V_{23} | \over 1 - |V_{13}|^{2}}  & - { |V_{11} V_{23}| \over 1 - |V_{13}|^{2}}  & |V_{33}| \\
\end{pmatrix} 
+ |V_{13}|
\begin{pmatrix}
 0 & 0 &  e^{- i \d} \\[2pt]
- { |V_{11} V_{23} | \over 1 - |V_{13}|^{2} } e^{ i\d} &- { |V_{12} V_{23}| \over 1 - |V_{13}|^{2} } e^{ i\d}   & 0 \\[2pt]
- { |V_{11} V_{33}| \over 1 - |V_{13}|^{2} } e^{ i\d} &- { |V_{12} V_{33}| \over 1 - |V_{13}|^{2} }  e^{ i\d}   & 0 \\
\end{pmatrix} . 
\end{align}
This decomposition directly reflects the intrinsic structure of CP phases in a unitary matrix. 
The even-order terms of $V_{13}$ depend only on relative phases, 
while the odd-order terms represent contributions that depend on the genuine CP phases $\delta$.
By this formulation, physical quantities described in the PDG parametrization can be directly translated into general rephasing invariants.

In this reduction, the independent variables appear to be $V_{11}, V_{12}, V_{13}, V_{23}, V_{33}$, and $\det V$, which  imply twelve degrees of freedom. However, evaluating the determinant,
\begin{align}
\det V = 
\frac{ ( | V_{11}|^2+ |V_{12}|^2 ) ( |V_{23} |^2 + | V_{33}|^2 )}
{ ( | V_{13}|^2 - 1)^2 \det V^*} \, . 
\end{align}
Taking into account the following three constraints, 
\begin{align}
 | V_{11}|^2+ |V_{12}|^2 + | V_{13}|^2  =  | V_{13}|^2 + |V_{23} |^2 + | V_{33}|^2 = \det V^* \det V = 1 \, , 
\end{align}
the number of independent degrees of freedom is reduced to nine, which is consistent. 
In other words, eliminating three additional absolute values from Eq.~(\ref{reduction})—by rewriting the determinant as $\det V = e^{i \arg \det V}$—removes all redundant variables arising from the unitarity. The result is a complete reduction expressed in terms of nine parameters and independent of any particular basis or parametrization.
To summarize, by using the inversion formula partially, a unitary matrix in an arbitrary basis is decomposed into the basis independent reduction of the unitarity constraints and the explicit rephasing transformation.

Furthermore, this reduction provides insight into the perturbative behavior associated with finite $U_{13}^{f}$. Restricting attention to terms up to first order of $|V_{13}| \ll 1$ in the rephasing invariant representation,
\begin{align}
V^{0} & \simeq 
\begin{pmatrix}
|V_{11}| & |V_{12}| & 0\\[2pt]
-  |V_{12} V_{33}  &  |V_{11} V_{33} |  & |V_{23}| \\[2pt]
 |V_{12} V_{23} |   & -  |V_{11} V_{23}|  & |V_{33}| \\
\end{pmatrix} 
+ |V_{13}|
\begin{pmatrix}
 0 & 0 &  e^{- i \d} \\[2pt]
-  |V_{11} V_{23} |  e^{ i\d} &-  |V_{12} V_{23}| e^{ i\d}   & 0 \\[2pt]
-  |V_{11} V_{33}| e^{ i\d} &-  |V_{12} V_{33}|   e^{ i\d}   & 0 \\
\end{pmatrix} .
\end{align}
This corresponds to the limit $c_{13} \to 1$ in the PDG parametrization. 
Considering such a differential structure is useful when analyzing the effects of a finite $U_{13}^{e}$.

\subsection{The limit of $U_{23}^{e} = 0$}

Motivated by the behavior of the CKM matrix and grand unified theories, 
it seems more appropriate to treat  $U_{12}^{e}$ exactly and $U_{23}^{e}$ perturbatively in the general expression of $\d$~(\ref{generald}). 
However, performing the approximation in this order inevitably involves nontrivial arguments, 
rendering the expression cumbersome.

In practice, imposing $U_{23}^{e} = 0$,
\begin{align}
U &= 
\begin{pmatrix}
U_{11}^{e *} & - {U_{12}^e U_{33}^e \over \det U^{e} } & 0  \\[2pt]
U_{12}^{e *} & {U_{11}^e U_{33}^e \over \det U^{e} } & 0 \\[2pt]
0 & 0 & U_{33}^{e *} \\
\end{pmatrix} 
\begin{pmatrix}
U_{11}^{\n} & U_{12}^{\n} & U_{13}^{\n} \\[2pt]
U_{21}^{\n} & U_{22}^{\n} & U_{23}^{\n} \\[2pt]
U_{31}^{\n} & U_{32}^{\n} & U_{33}^{\n} \\
\end{pmatrix} \nn
\\ & = 
\begin{pmatrix}
U_{11}^{e *} U_{11}^{\n} - {U_{12}^e U_{33}^e \over \det U^{e} } U_{21}^{\n} & 
U_{11}^{e *} U_{12}^{\n} - {U_{12}^e U_{33}^e \over \det U^{e} } U_{22}^{\n} & 
U_{11}^{e *} U_{13}^{\n} - {U_{12}^e U_{33}^e \over \det U^{e} } U_{23}^{\n} \\[2pt]
*  &  * & U_{12}^{e *}  U_{13}^{\n} +  {U_{11}^e \over \det U^{e} } U_{33}^e U_{23}^{\n}   \\[2pt]
* & * &  U_{33}^{e *}U_{33}^{\n}
\end{pmatrix} . 
\end{align}
The CP phase of this mixing matrix is
\begin{align}
\d = \arg\left[  
{ U_{11}^{e *} U_{11}^{\n} - {U_{12}^e U_{33}^e \over \det U^{e} } U_{21}^{\n} 
\over (U_{11}^{e *} U_{12}^{\n} - {U_{12}^e U_{33}^e \over \det U^{e} } U_{22}^{\n} )^{*}}
{U_{12}^{e *}  U_{13}^{\n} +  {U_{11}^e U_{33}^e \over \det U^{e} }  U_{23}^{\n} 
\over U_{11}^{e *} U_{13}^{\n} - {U_{12}^e U_{33}^e \over \det U^{e} } U_{23}^{\n}  }
{ U_{33}^{e *}U_{33}^{\n} \over \det U^{e *} \det U^{\n} }
\right ] . 
\end{align}
Note that this CP phase reduces to $\d^{\n}$ in the further limit $U_{12}^{e} \to 0$.

In principle, this phase should be a function of $\d^{\n}$ and $\r_{1}-\r_{2}$.
To see this explicitly, the matrix elements $U_{21}^{\n}$ and $U_{22}^{\n}$,
which carry nontrivial phases in $U_{e1}$ and $U_{e2}$, must be expressed in terms of the remaining elements.

Starting from the relevant phase $\arg[U_{e1} / U_{e2}^{*}]$,
one can eliminate $U_{21}^{\n}$ and $U_{22}^{\n}$ by using the reduction~(\ref{reduction}) 
\begin{align}
\arg \left [ {U_{e1} \over U_{e2}^{*}} \right ] - \arg \left [ {U_{11}^{\n} \over U_{12}^{\n *}} \right ]  & = 
\arg\left[  
 \lsp U_{11}^{e *}  - {U_{12}^e U_{33}^e \over \det U^{e} } { U_{21}^{\n} \over U_{11}^{\n} } \rsp
\bigg / \lsp U_{11}^{e *}  - {U_{12}^e U_{33}^e \over \det U^{e} } {U_{22}^{\n} \over U_{12}^{\n} } \rsp^{*}  \right ] \nn  \\
& =
\arg\left[  
{  U_{11}^{e *} + {U_{12}^e U_{33}^e \over \det U^{e} ( 1 - |U^{\n}_{13}|^{2})} 
   ( {U_{12}^{\n *} U_{33}^{\n *}  \over \det U^{\n *} U_{11}^{\n} } + U_{13}^{\n *}  U^{\n}_{23}  ) \over 
 \lsp U_{11}^{e *}  - {U_{12}^e U_{33}^e \over \det U^{e} (1 - |U^{\n}_{13}|^{2}) }
  ( {U_{11}^{\n *} U_{33}^{\n *}  \over \det U^{\n *} U_{12}^{\n}  } - U_{13}^{\n *}  U^{\n}_{23}  ) \rsp^{*} } \right ] .
\end{align}
Apart from the overall phase $\arg [U_{11}^{e *} / U_{11}^{e}]$, this phase depends on the following two complex quantities whose arguments are $\arg R_{12} = \r_{1} - \r_{2}$ and $\arg D^{\n} = \d^{\n}$ 
\begin{align}
R_{12} \equiv { U^{e *}_{11}  U^{e *}_{12}  U^{e *}_{33} \over \det U^{e *}  } { U^{\n}_{11} U^{\n}_{12}  U^{\n}_{33} \over \det U^{\n} } \, ,  ~~
D^{\n} \equiv  {U_{11}^{\n } U_{12}^{\n } U_{23}^{\n } U_{3 3}^{\n }   \over U_{13}^{\n } \det U^{\n } } \, , 
~~
{R_{12} \over D^{\n} } = 
{ U^{e *}_{11}  U^{e *}_{12}  U^{e *}_{33} \over \det U^{e *}  } { U^{\n}_{13} \over U^{\n}_{23}  } \, . 
\end{align}
The remaining part of $\d$ carries no nontrivial phase as
\begin{align}
\arg \left [ {U_{\m 3} U_{\t 3} \over U_{e3} \det U} \right ] - \arg \left [ {U_{23}^{\n} U^{\n}_{33}\over U_{13}^{\n} \det U^{\n} } \right ]  & = 
\arg\left[  
{  {U_{11}^e U_{33}^e \over \det U^{e} }  + U_{12}^{e *}  { U_{13}^{\n} \over U_{23}^{\n}}   
\over U_{11}^{e *} - {U_{12}^e U_{33}^e \over   \det U^{e} } {U_{23}^{\n} \over  U_{13}^{\n}} }
{ U_{33}^{e *}  \over \det U^{e *} }
\right ] 
= 
\arg\left[  
 {U_{11}^e |U_{33}^e|^{2}  + {U_{12}^{e *} U_{33}^{e *} \over \det U^{e *} } { U_{13}^{\n} \over U_{23}^{\n}}   
\over U_{11}^{e *} - {U_{12}^e U_{33}^e \over   \det U^{e} } {U_{23}^{\n} \over  U_{13}^{\n}} }
\right ] . 
\end{align}
By appropriately adding $\arg U_{11}^{e}$ and $\arg U^{e *}_{11}$, 
the phase $\d$ is entirely expressed in terms of rephasing invariants, 
and it indeed depends on $\d^{\n}$ and $\r_{1} - \r_{2}$; 
\begin{align}
\d &= \d^{\n} + \arg\left[  
{  |U_{11}^{e}|^{2} + {U_{11}^{e} U_{12}^e U_{33}^e \over \det U^{e} ( 1 - |U^{\n}_{13}|^{2})} 
   ( {U_{12}^{\n *} U_{33}^{\n *}  \over \det U^{\n *} U_{11}^{\n} } + U_{13}^{\n *}  U^{\n}_{23}  ) \over 
 \lsp |U_{11}^{e}|^{2}  - {U_{11}^{e} U_{12}^e U_{33}^e \over \det U^{e} (1 - |U^{\n}_{13}|^{2}) }
  ( {U_{11}^{\n *} U_{33}^{\n *}  \over \det U^{\n *} U_{12}^{\n}  } - U_{13}^{\n *}  U^{\n}_{23}  ) \rsp^{*} } \right ] 
  + \arg\left[  
 { |U_{11}^e U_{33}^e|^{2}  + {U_{11}^{e *} U_{12}^{e *} U_{33}^{e *} \over \det U^{e *} } { U_{13}^{\n} \over U_{23}^{\n}}   
\over |U_{11}^{e}|^{2} - {U_{11}^{e} U_{12}^e U_{33}^e \over \det U^{e} } {U_{23}^{\n} \over  U_{13}^{\n}} }
\right ] \nn \\
& = \d^{\n} + \arg\left[  
{  |U_{11}^{e}|^{2} + {R_{12}^{*}  \over 1 - |U^{\n}_{13}|^{2} } 
   ( { 1 \over |U_{11}^{\n}|^{2} } + {|U_{23}^{\n}|^{2}\over D^{\n*} }  ) \over 
 |U_{11}^{e}|^{2}  - { R_{12} \over 1 - |U^{\n}_{13}|^{2} }
  ( {1 \over |U_{12}^{\n}|^{2} } - {|U_{23}^{\n}|^{2}\over D^{\n} }   ) } \right ] 
  + \arg\left[  
 { |U_{11}^e U_{33}^e|^{2}  + (R_{12} / D_{\n})
\over |U_{11}^{e}|^{2} - {|U_{23}^{\n}|^{2} R_{12}^{*} \over  |U_{13}^{\n}|^{2} D_{\n}^{*} } }
\right ] . 
\end{align}

Although one can also write it using a suitable parametrization as in the simplified case above, 
the resulting form is not particularly instructive. 
For treating finite $U_{12}^{e}$ and perturbatively small $U_{23}^{e}$, the alternative parametrization proposed by Fritzsch--Xing can be more appropriate \cite{Fritzsch:1997fw}, 
because the nontrivial phases appear in the perturbative sector \cite{Yang:2026wjg}. 

\section{Summary}

In this paper, we explore rephasing invariant structures of the Dirac CP phase $\d$ under an approximation $U^{e}_{13} = 0$, where the 1-3 element of the diagonalization of charged leptons $U^{e}$ is neglected. 
With the further simplified condition $U^{e}_{12} = 0$, the Dirac phase reduces to a compact form 
$\delta =  \delta^{\nu} + \arg [ (U_{33}^e / U_{23}^{e}) - ( U_{33}^{\n} /  U_{23}^{\n}) ] 
- \arg [ ( U_{33}^e / U_{23}^{e}) + (U_{23}^{\n *} / U_{33}^{\n*} ) ]  $, 
and the CP phase for finite $U_{12}^{e}$ can be understood as a generalization of this compact form. 
These results encompass almost all perturbative calculations of the CP phases
in quark and lepton mixing matrices with hierarchical masses of charged fermions.

As a second result of this work, we derive a basis independent reduction of the unitarity constraints for an arbitrary unitary matrix $V$ by eliminating the elements $V_{21}, V_{22}, V_{31}, V_{32}$ using the inversion formula.
This reduction is expressed solely in terms of $V_{11}, V_{12}, V_{13}, V_{23}, V_{33}$ and $\det V$. 
By further imposing the three constraints $\sum_{j} |V_{1j}|^{2} = \sum_{i} |V_{i 3}| = |\det V| = 1$, 
all redundancy is removed, and any unitary matrix in an arbitrary basis is represented explicitly with nine parameters.
Applying the explicit rephasing transformation to this reduction yields a rephasing invariant representation of the PDG parametrization containing the CP phase $\delta$. This representation allows the translation of theoretical results expressed in the PDG parametrization directly into the language of rephasing invariants.

These results are independent of any specific parametrization and thus constitute a useful general tool 
in analysis of the CP symmetry and generalized CP transformations.
This rephasing invariant formulation clarifies structures of CP phases in both quarks and leptons, 
and provides a systematic framework for understanding their role in flavor symmetries and grand unified theories.

\section*{Acknowledgment}

The study is partly supported by the MEXT Leading Initiative for Excellent Young Researchers Grant Number JP2023L0013.


\end{document}